\documentclass[aps,pre,amsmath,amssymb,twocolumn]{revtex4}
\usepackage{graphicx}
\usepackage{epstopdf}
\usepackage{pstricks}
\usepackage{xcolor}
\usepackage{subcaption}
\newcommand{\be}{\begin{equation}}
\newcommand{\ee}{\end{equation}}
\newcommand{\ba}{\begin{array}{l}}
\newcommand{\ea}{\end{array}}
\newcommand{\re}[1]{(\ref{#1})}
\newcommand{\ci}[1]{\cite{#1}}
\newcommand{\banonum}{\begin{eqnarray*}}
\newcommand{\eanonum}{\end{eqnarray*}}
\newcommand{\baa}{\begin{eqnarray}}
\newcommand{\eaa}{\end{eqnarray}}
\newcommand{\bfr}{\begin{flushright}}
\newcommand{\efr}{\end{flushright}}
\newcommand{\bfl}{\begin{flushleft}}
\newcommand{\efl}{\end{flushleft}}
\newcommand{\lab}[1]{\label{#1}}

\begin{document}
 \title{Nonlocal Nonlinear Schrodinger Equation on Metric Graphs}
\author{ K. Sabirov$^a$, D. Matrasulov$^{b,c}$, M. Akramov$^{b,c}$, H. Susanto$^d$}

\affiliation{
$^a$Tashkent University of  Information Technology, Amir Temur Avenue 108, Tashkent 100200, Uzbekistan\\
${^b}$Turin Polytechnic University in Tashkent, 17 Niyazov Str.,
100095,  Tashkent, Uzbekistan\\
${^c}$Yeoju Technical Institute in Tashkent, 156 Usman Nasyr Str., 100121, Tashkent, Uzbekistan\\
 $^d$Department of Mathematics, Khalifa University, Abu Dhabi Campus, PO Box 127788, United Arab Emirates}

\begin{abstract}
 We consider PT-symmetric, nonlocal nonlinear Schrodinger equation on metric
 graphs. Vertex boundary conditions are derived from the conservation laws. 
 Soliton solutions are obtained for simplest graph topologies, such as star and tree graphs.  Integrability of the problem is shown by proving existence of infinite number of conservation laws. 

\end{abstract}

\maketitle

\section{Introduction}

PT-symmetric nonlocal nonlinear Schrodinger (NNLS) equation has attracted much
attention since from the pioneering paper by Ablowitz and
Muslimani \ci{AM2013}, where the soliton solutions are obtained using inverse scattering based approach. Remarkable feature of the problem is its integrability, which was shown \ci{AM2013}. Different aspects of nonlinear nonlocal Schrodinger equation, such as integrability, various soliton solutions and their properties have been studied during past few years \ci{AM2013}-\ci{Panos2020}. In \ci{AM2014}
discrete  version of NNLS equation have been
considered and its integrability was shown. In \ci{AM2016}
extended analysis of  NNLS equation, which includes details of the inverse
scattering, Riemann-Hilbert  and Cauchy problems, was presented.
In \ci{Sinha} exact solutions of different versions of NNLS equation have
been obtained. Ref.\ci{Yang} presents study of a physically
significant version of NNLSE which can be derived from the Manakov
system. General soliton solution of  a nonlocal nonlinear
Schrodinger equation with zero and nonzero boundary conditions was
derived in \ci{AM2018}. Quasi-monochromatic complex reductions of
a cubic nonlinear Klein-Gordon, the KdV and water waves equations
and their relations to nonlocal PT-symmetric nonlinear Schrodinger
equation was studied in the Ref.\ci{AM2019}. Rogue waves and
periodic solutions in an  NNLS equation based model have been studied in
the recent Ref.\ci{Panos2020}.  In this paper we consider extension of
the  Ablowitz-Muslimani NNLS equation to the case of branched 1D
domains called the metric graphs. These latter are the
one-dimensional wires (bonds) connected to each other according to
some rule, which is called topology of a graph. Each bond is
assumed to assigned a length. Motivation for the study of NNLS equation on
graphs comes from the fact that it describes nonlocal solitons in
branched optical materials providing self-induced gain-and-loss.
In such materials branching topology can be used for tuning of the fiber's properties  and controlling the optical signal transfer. In the next
section we briefly recall NNLS equation on a line, following the
Ref.\ci{AM2013}. Section III presents formulation of the problem
and its soliton solutions for a star branched graph. In section IV we provide proof for the integrability of NNLS equation on a star graph. In section V we present numerical
results on the dynamics of nonlocal solitons on metric star graph. Section VI considers NNLS equation for a tree graph. Finally, section V presents some concluding remarks.

\begin{figure}[t!]
\includegraphics[width=80mm]{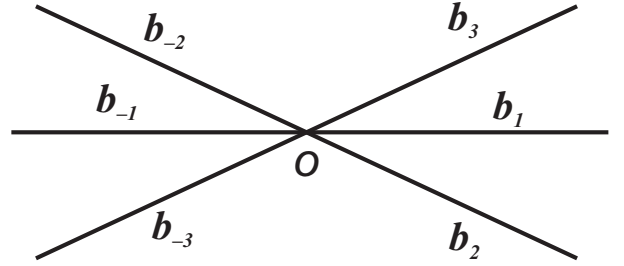}
\caption{Star graph with six bonds.} \label{pic1}
\end{figure}

\section{Nonlocal nonlinear Schrodinger equation on a line}

PT-symmetric version of the nonlocal nonlinear Schrodinger
equation on a line was proposed by Ablowitz and
Muslimani in \ci{AM2013} and was studied later in different contexts, e.g., 
 Refs.\ci{AM2014} -\ci{Panos2020}. Explicitly, NNLSE on a line can be written as
\begin{equation}
i\frac{\partial}{\partial t}q(x,t)=\frac{\partial^2}{\partial
x^2}q(x,t)+2 q^2(x,t)q^*(-x,t).\label{nlse0}
\end{equation}

Eq.\re{nlse0} can be rewritten as
\begin{equation}
\frac{\partial}{\partial t}q(x,t)=-i\frac{\partial^2}{\partial
x^2}q(x,t)+iV(x,t),\label{nlse01}
\end{equation}
where $V=-2q^2(x,t)q^*(-x,t)$ is the PT-symmetric self-induced
potential. Eq.\re{nlse01} describes the PT-symmetric optical solitons 
propagating in optical waveguide having "gain-and-loss" structure.
One-soliton solution of Eq.\re{nlse0} can be obtained using
the inverse scattering method and given as \ci{AM2013}
\be
q(x,t)=-\frac{2(\eta_1+\bar{\eta}_1)e^{i\bar{\theta}_1}e^{-4i\bar{\eta}^2_1t}e^{-2\bar{\eta}_1x}}{1+e^{i(\theta_1+\bar{\theta}_1)}+e^{4i(\eta^2_1-\bar{\eta}^2_1)t}+e^{-2(\eta_1+\bar{\eta}_1)x}}.
\lab{sol01} \ee

As it was stated in the Ref.\ci{AM2013}, solution given by
Eq.\re{sol01} describes breathing soliton, whose center of mass
oscillates around a fixed point.
A bright (traveling) soliton solution was found in the
Ref\ci{Stalin} and can be written as
\begin{eqnarray}
q(x,t)=\frac{\alpha
e^{\bar{\xi}_1}}{1+e^{\xi_1+\bar{\xi}_1+\Delta}},\label{sol2}
\end{eqnarray}
and the parity transformed complex conjugate solution
\begin{eqnarray}
q^*(-x,t)=\frac{\beta
e^{\xi_1}}{1+e^{\xi_1+\bar{\xi}_1+\Delta}},\label{sol3}
\end{eqnarray}
where
$\xi_1=ik_1x-ik_1^2t+\xi_1^{(0)},\,\bar{\xi}_1=i\bar{k}_1x-i\bar{k}_1^2t+\bar{\xi}_1^{(0)},\,e^{\Delta}=-\frac{\alpha\beta}{\kappa},\,\kappa=(k_1+\bar{k}_1)^2$
and $k_1,\bar{k}_1,\alpha,\beta,\xi_1^{(0)},\bar{\xi}_1^{(0)}$ are
arbitrary complex constants.
Two conservative quantities (integrals of motion) can be
determined for Eq.\re{nlse0} as the norm:
\begin{eqnarray}
C_0(t)=\underset{-\infty}{\overset{+\infty}{\int}}q(x,t)q^*(-x,t)dx.\label{norm1}
\end{eqnarray}
and energy:
\begin{gather}
C_2(t)=\underset{-\infty}{\overset{+\infty}{\int}}\Big[\frac{\partial}{\partial
x}q(x,t)\cdot\frac{\partial}{\partial
x}q^*(-x,t)+\nonumber\\
q^2(x,t)\cdot
q^{*2}(-x,t)\Big]dx.\label{energy1}
\end{gather}
Integrability of Eq.\re{nlse0} was
proven  in the Ref.\ci{AM2013} by showing existence of the
infinite number of conserving quantities. In the next section we
will extend the study of the Ref.\ci{AM2013} to the case one dimensional branched domains given in terms of the metric graphs.

\section{Extension to  a star graph: Vertex boundary conditions and soliton solutions}
Consider the following nonlocal nonlinear Schr\"odinger equation
which is written  on the each bond of the  star graph with six
semi-infinite bonds (see, Fig. 1)
$b_{-j}\sim(-\infty;0]$ and $b_{j}\sim[0;+\infty),$
\begin{gather}
i\frac{\partial}{\partial t} q_{\pm j}(x,t)=\frac{\partial^2}{\partial
x^2}q_{\pm j}(x,t)+\sqrt{\beta_{j}\beta_{-j}}q_{\pm j}^2(x,t)q_{\mp j}^* (-x,t),\label{nlse1}
\end{gather}
where $q_{\pm j}(x,t)$ at $x\in b_{\pm j}$ and $j=1,2,3$.

\textit{\bf Very important feature of Eq.\re{nlse1} is the fact that it is a system of NLS equations, where components of $q_j$ are mixed in nonlinear term. Unlike classical NLSE on graphs, where components of solution are related to each other via the vertex boundary conditions, in Eq.\re{nlse1}, components with opposite signs are mixed via the nonlinear term, while other components are connected to each other via the vertex boundary conditions.}
To consider and solve Eq.\re{nlse1}, one needs to impose boundary
conditions at the graph branching point (vertex). Such conditions
can be derived from the fundamental conservation laws. Here we use
norm and energy conservation to derive vertex boundary conditions.

For the above nonlocal NLSE, the norm determined as \ci{AM2013}
\begin{eqnarray}
C_0(t)=\underset{j=1}{\overset{3}{\sum}}  \bigg[
\underset{b_{j}}{\int}q_{j}(x,t)q_{-j}^*(-x,t)dx+\nonumber\\
\underset{b_{-j}}{\int}q_{-j}(x,t)q_{j}^*(-x,t)dx
\bigg].\label{norm1}
\end{eqnarray}
From the norm conservation, $\dot{C}_0=0$ we have
\begin{eqnarray}
\sum_{j=1}^3\left.{\bf Im}\left[\frac{\partial}{\partial x}q_{j}(x,t)\cdot
q_{-j}^*(-x,t)\right]\right|_{x\to+0}=\nonumber\\
\sum_{j=1}^3\left.{\bf Im}\left[\frac{\partial}{\partial x}q_{-j}(x,t)\cdot
q_{j}^*(-x,t)\right]\right|_{x\to-0}.\label{norm3}
\end{eqnarray}

Another conserving quantity, i.e., the energy is given by
\begin{widetext}
\begin{eqnarray}
C_2(t)=\underset{j=1}{\overset{3}{\sum}} 
\Bigg[
\underset{b_{j}}{\int} 
\left(\frac{\partial}{\partial
x}q_{j}(x,t)\cdot\frac{\partial}{\partial
x}q_{-j}^*(-x,t)-\frac{\sqrt{\beta_{j} \beta_{-j} }}{2} q_{j}^2(x,t)\cdot q_{-j}^{*2}(-x,t)\right) dx+\nonumber\\
\underset{b_{-j}}{\int}
\left(\frac{\partial}{\partial
x}q_{-j}(x,t)\cdot\frac{\partial}{\partial
x}q_{j}^*(-x,t)-\frac{\sqrt{\beta_{j}\beta_{-j}}}{2} q_{-j}^2(x,t)\cdot q_{j}^{*2}(-x,t)\right) dx \Bigg].  \label{energy1}
\end{eqnarray}
\end{widetext}

The energy conservation, $\dot{C}_2=0$
leads to
\begin{eqnarray}
\sum_{j=1}^3 \left.{\bf Re}\left[\frac{\partial}{\partial t}q_{-j}^*(-x,t)\cdot
\frac{\partial}{\partial
x}q_{j}(x,t)\right]\right|_{x\to+0}=\nonumber\\
\sum_{j=1}^3 \left.{\bf Re}\left[\frac{\partial}{\partial t}q_{j}^*(-x,t)\cdot
\frac{\partial}{\partial
x}q_{-j}(x,t)\right]\right|_{x\to-0}.\label{energy2}
\end{eqnarray}
Eqs. (\ref{norm3}) and (\ref{energy2}) are compatible with the
following two sets of the vertex boundary conditions:
\begin{widetext}
\begin{gather}
\alpha_{1} q_{1}(x,t)|_{x=0}=\alpha_{-1}q_{-1}(x,t)|_{x=0}=\alpha_{2}q_{2}(x,t)|_{x=0}=\alpha_{-2} q_{-2}(x,t)|_{x=0}=\alpha_{3} q_{3}(x,t)|_{x=0}=\alpha_{-3} q_{-3}(x,t)|_{x=0},\nonumber\\
\left.\frac{1}{\alpha_{1}}\frac{\partial}{\partial
x}q_{1}(x,t)\right|_{x=0}+
\left.\frac{1}{\alpha_{2}}\frac{\partial}{\partial
x}q_{2}(x,t)\right|_{x=0}+
\left.\frac{1}{\alpha_{3}}\frac{\partial}{\partial
x}q_{3}(x,t)\right|_{x=0}=\nonumber\\
\left.\frac{1}{\alpha_{-1}}\frac{\partial}{\partial
x}q_{-1}(x,t)\right|_{x=0}+
\left.\frac{1}{\alpha_{-2}}\frac{\partial}{\partial
x}q_{-2}(x,t)\right|_{x=0}+
\left.\frac{1}{\alpha_{-3}}\frac{\partial}{\partial
x}q_{-3}(x,t)\right|_{x=0}
.\label{bc1}
\end{gather}
\end{widetext}

It should be noted that Eqs. (\ref{norm3}) and (\ref{energy2})
follow from the boundary conditions \re{bc1}, but
vice-versa is not true.
Let $q(x,t)$ is the solution of nonlocal nonlinear Schrodinger
equation given by
\begin{equation}
i\frac{\partial}{\partial t}q(x,t)=\frac{\partial^2}{\partial
x^2}q(x,t)+2q^2(x,t)q^*(-x,t).\label{nlse2}
\end{equation}
Then solution of  Eq.  (\ref{nlse1}) and \re{bc1} can be expressed
in terms of $q(x,t)$ as $q_{\pm j}(x,t)=\sqrt{\frac{2}{\beta_{\pm j}}}q(x,t)$
and fulfills the boundary conditions  \re{bc1}, provided the
following constraints hold true:
\begin{gather}
\frac{\alpha_{\pm j}}{\alpha_{1}}=\sqrt{\frac{\beta_{\pm j}}{\beta_{1}}},\nonumber\\
\frac{1}{\beta_{1}}+\frac{1}{\beta_{2}}+\frac{1}{\beta_{3}}=
\frac{1}{\beta_{-1}}+\frac{1}{\beta_{-2}}+\frac{1}{\beta_{-3}}.\label{constrain1}
\end{gather}
One of explicit soliton solutions of Eq. (\ref{nlse1}) on a line
have been obtained in the Ref.  is \cite{AM2016}. Using this
solution, corresponding soliton solution on a graph can be written
as
\begin{eqnarray}
q_{\pm j}(x,t)=-\sqrt{\frac{2}{\beta_{\pm j}}}\frac{4\eta
e^{i\bar{\varphi}}e^{-4i\eta^2t}e^{-2\eta
x}}{1+e^{i(\varphi+\bar{\varphi})}e^{-4\eta x}}.\label{sol1}
\end{eqnarray}
$\varphi,\,\bar{\varphi},\,\eta$ are arbitrary complex constants.

\section{Integrability of the problem}
Here we will show integrability of the NNLSE  on a metric star
graph, given by Eqs.\re{nlse1} and \re{bc1} by proving
existence of the infinite number of conservation laws. This can be
done following the prescription used for usual (not nonlocal) NLS
on metric graphs in the Ref. \ci{Zarif}. The soliton solutions of
the problem on the infinite linear chain satisfy an infinite
number of conservation laws given by \be
\underset{-\infty}{\overset{+\infty}{\int}}\mu_n[q(x,t),q^*(-x,t)]dx=C_n
\lab{consl},\ee where $C_n$ is a constant, $\mu_n$ is a polynomial
of $q(x,t),\,q^*(-x,t)$ and their derivatives with respect to $x$
\cite{AM2016}. Using this relation, we now investigate the
following quantities, which for are given on the metric star
graph:
\begin{eqnarray}
Q_n=\underset{j=1}{\overset{3}{\sum}}\bigg[ \beta_j^{-1}\underset{b_{j}}{\int}\mu_n[q(x,t),q^*(-x,t)]dx+\nonumber\\
\beta_{-j}^{-1}\underset{b_{-j}}{\int}\mu_n[q(x,t),q^*(-x,t)]dx\bigg],\label{claws1}\label{claws1}
\end{eqnarray}  
where $q(x,t)$ the solution of Eq. (\ref{nlse1}) in the bonds
$b_{\pm j}$ and $\mu_n[q(x,t),q^*(-x,t)]$ obeys the
recursion relation
\begin{eqnarray}
&\mu_{n+1}=q\frac{\partial}{\partial x}\left(\frac{\mu_n}{q}\right)+\underset{m=0}{\overset{n-1}{\sum}}\mu_m\mu_{n-m-1},\label{claws2}\\
&\mu_0=q(x,t)q^*(-x,t),\,\mu_1=q(x,t)\partial_xq^*(-x,t).\label{claws3}
\end{eqnarray}
Using Eq. (\ref{constrain1}), from the r.h.s. of Eq.(\ref{claws1})
we can get
\begin{gather}
Q_n=(\beta_{-1}^{-1}+\beta_{-2}^{-1}+\beta_{-3}^{-1})
\underset{-\infty}{\overset{0}{\int}}\mu_n[q(x,t),q^*(-x,t)]dx+\nonumber\\
(\beta_{1}^{-1}+\beta_{2}^{-1}+\beta_{3}^{-1})\underset{0}{\overset{+\infty}{\int}}\mu_n[q(x,t),q^*(-x,t)]dx=\nonumber\\
=(\beta_{1}^{-1}+\beta_{2}^{-1}+\beta_{3}^{-1})\underset{-\infty}{\overset{+\infty}{\int}}\mu_n[q(x,t),q^*(-x,t)]dx=
\nonumber\\
(\beta_{1}^{-1}+\beta_{2}^{-1}+\beta_{3}^{-1})C_n.\label{claws4}
\end{gather}
It is clear that due to the conservation law given by
Eq.\re{consl}, $Q_n$ is constant, i.e.
conserving quantity. Therefore,  $Q_n$ is
the constant of motion. This implies that nonlocal NLSE on metric
star graph has infinite number of conservation laws and hence,
integrable.

\section{Numerical results}
It is clear that the above proven integrability of NNLSE
\eqref{nlse1} holds true for the case, when constraints given by
Eq.\re{constrain1} are fulfilled. Such integrable NLSE approves
different soliton solutions, such as breathing given by
Eq.\re{sol01} and traveling given by Eq.\re{sol2} solitons. For
the case, when constraints in Eq.\re{constrain1} are broken, one
needs to solve Eq. \eqref{nlse1}  numerically as the initial
value problem. As the initial conditions, we will choose
values of exact (soliton) solutions at $t=0$. Discretization scheme from the Ref.\ci{AM2014} is used in numerical solutioin of Eq.\re{nlse0}.  In Fig. 2 
plots of $|q(x,t)|^2$ obtained by solving Eq.\eqref{nlse1}
numerically for the initial conditions imposed on the bond $b_{-1}$ and $b_1$ are presented for different time moments, $t=0\;\;0.05,\;\;0.1$ at the values of $\beta_{\pm j}$ fulfilling the sum rule in Eq. \eqref{constrain1}. A remarkable feature of the travelling solitons is the reflectionless transmission through the vertex. Fig. 3 presents similar plots for those values of $\beta_{\pm j}$, which do not fulfill the sum rule in \eqref{constrain1}. Unlike the solitons in Fig. 2, reflection at the vertex can be observed in this plot. Thus one can conclude that integrable case provides the reflectionless transmission of solitons through the branching point of the graph. |Earlier, such a feature was observed for other evolution equations on graphs, such as NLS \cite{Zarif}, sine-Gordon \cite{Our1} and nonlinear Dirac \cite{KarimNLDE} equations. The reason for such behavior of solitons described by the nonlinear Schrodinger equation on graphs, was explained in the Ref.\cite{Jambul1}.

\begin{figure}[t!]
\includegraphics[width=90mm]{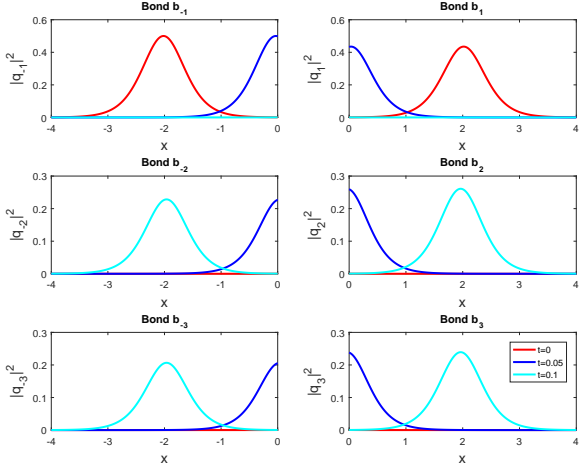}
\caption{Plot of soliton on a metric star graph obtained from numerical solution of Eq.\eqref{nlse1} for the values of $\beta_j$ fulfilling the sum rule in Eq.\eqref{nlse1} ($\beta_{-1}=1$, $\beta_1=1.15$, $\beta_{-2}=2.19$, $\beta_2=1.91$, $\beta_{-3}=2.42$, $\beta_3=2.09$). The initial conditions are given on the bonds $b_{-1}$ and $b_1$.} \label{pic2}
\end{figure}

\begin{figure}[t!]
\includegraphics[width=90mm]{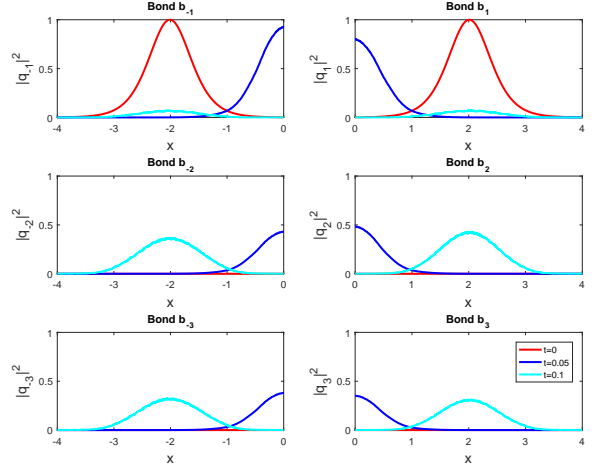}
\caption{Plot of soliton on a metric star graph obtained from numerical solution of Eq.\eqref{nlse1} for the values of $\beta_j$ breaking the sum rule in Eq.\eqref{nlse1} ($\beta_{-1}=0.65$, $\beta_1=0.79$, $\beta_{-2}=2.7$, $\beta_2=2.09$, $\beta_{-3}=3.06$, $\beta_3=2.87$). The initial conditions are given on the bonds $b_{-1}$ and $b_1$.} \label{pic3}
\end{figure}

\section{Extension to a tree graph}
The above treatment can be extended to the case other graphs. Here we will demonstrate that for a tree graph. One of possible tree graphs, on which one can write NNLSE, is presented in Fig.4. The central branch, i.e. the branch st the middle of the graph is chosen as an origin of coordinates. Then the bonds can be determined as   
$b_{-1}, \; b_{-1mn}\sim (-\infty; 0]$,
$b_{-1m}\sim [-L_{1m}; 0]$, 
$b_{1m}\sim [0; L_{1m}]$,
$b_{1},\; b_{1mn}\sim [0; +\infty)$, where $L_{1m}$ are the lengths of $b_{\pm 1m}$ bonds and $m=1,2$, $n=1,2$. Here the "+" sign is for right-handed bonds and the "-" sign is for left-handed bonds from the center of the tree graph.

On each bond of such graph one can write nonlocal nonlinear Schrodinger equation given be Eq.\eqref{nlse1} with $j=\pm 1,\; \pm 1m,\;  \pm 1mn$.

The vertex boundary conditions following from the conservation laws are given as
\begin{gather}
\alpha_{\pm 1}q_{\pm 1}(x,t)|_{x=0}=
\alpha_{\mp 11}q_{\mp 11}(x,t)|_{x=0} =
\alpha_{\mp 12} q_{\mp 12}(x,t)|_{x=0},\nonumber\\
\left.\frac{1}{\alpha_{\pm 1}}\frac{\partial}{\partial
x}q_{\pm 1}(x,t)\right|_{x=0}=
\left.\frac{1}{\alpha_{\mp 11}}\frac{\partial}{\partial
x}q_{\mp 11}(x,t)\right|_{x=0}+\nonumber\\
\left.\frac{1}{\alpha_{\mp 12}}\frac{\partial}{\partial
x}q_{\mp 12}(x,t)\right|_{x=0},
\end{gather}
\begin{gather}
\alpha_{\pm 1m}q_{\pm 1m}(x,t)|_{x=\pm L_{1m}}=
\alpha_{\pm 1m1} \lim_{x\to\pm 0} q_{\pm 1m1}(x,t)=\nonumber\\
\alpha_{\pm 1m2}\lim_{x\to\pm 0}q_{\pm 1m2}(x,t),\nonumber\\
\left.\frac{1}{\alpha_{\pm 1m}}\frac{\partial}{\partial
x}q_{\pm 1m}(x,t)\right|_{x=\pm L_{1m}}=
\frac{1}{\alpha_{\pm 1m1}}\lim_{x\to\pm 0}\frac{\partial}{\partial
x}q_{\pm 1m1}(x,t)+\nonumber\\
\frac{1}{\alpha_{\pm 1m2}}\lim_{x\to\pm 0}\frac{\partial}{\partial
x}q_{\pm 1m2}(x,t).
\end{gather}
Assuming that the following sum rules hold true:
\begin{gather}
\frac{\alpha_{\pm 1}}{\sqrt{\beta_{\pm 1}}}=\frac{\alpha_{\pm 11}}{\sqrt{\beta_{\pm 11}}}=\frac{\alpha_{\pm 12}}{\sqrt{\beta_{\pm 12}}}, \nonumber\\
\frac{1}{\alpha_{\pm 1}\sqrt{\beta_{\pm 1}}}=\frac{1}{\alpha_{\pm 11}\sqrt{\beta_{\pm 11}}}+\frac{1}{\alpha_{\pm 12}\sqrt{\beta_{\pm 12}}}, \nonumber\\
\frac{1}{\beta_{\pm 1}}=\frac{1}{\beta_{\pm 11}}+\frac{1}{\beta_{\pm 12}}.
\end{gather}

\begin{gather}
\frac{\alpha_{\pm 1m}}{\sqrt{\beta_{\pm 1m}}}=\frac{\alpha_{\pm 1m1}}{\sqrt{\beta_{\pm 1m1}}}=\frac{\alpha_{\pm 1m2}}{\sqrt{\beta_{\pm 1m2}}}, \nonumber\\ 
\frac{1}{\alpha_{\pm 1m}\sqrt{\beta_{\pm 1m}}}=\frac{1}{\alpha_{\pm 1m1}\sqrt{\beta_{\pm 1m1}}}+\frac{1}{\alpha_{\pm 1m2}\sqrt{\beta_{\pm 1m2}}}, \nonumber\\
\frac{1}{\beta_{\pm 1m}}=\frac{1}{\beta_{\pm 1m1}}+\frac{1}{\beta_{\pm 1m2}},
\end{gather}
the soliton solutions on each bond can be written as 
\begin{eqnarray}
q_{\pm 1}(x,t)=\sqrt{\frac{2}{\beta_{\pm 1}}}q(x+S_{\pm 1},t),\nonumber\\
q_{\pm 1m}(x,t)=\sqrt{\frac{2}{\beta_{\pm 1m}}}q(x+S_{\pm 1m},t),\nonumber\\
q_{\pm 1mn}(x,t)=\sqrt{\frac{2}{\beta_{\pm 1mn}}}q(x+S_{\pm 1mn},t),
\end{eqnarray}
\begin{figure}[t!]
\centering
\includegraphics[scale=0.5]{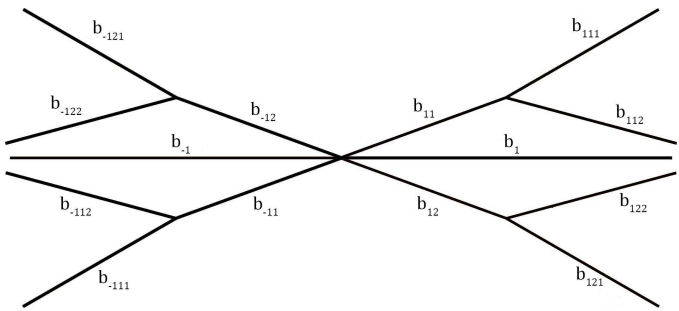}
\caption{A sketch of tree graph adopted for nonlocal nonlinear Schrodinger equation}
\end{figure}
where $S_{\pm 1}=S_{\pm 1m}=x_0$, $S_{\pm 1mn}=\pm L_{1m}+x_0$, $x_0$ is the coordinate of the center of  soliton at $t=0$. Integrability of nonlocal nonlinear Schrodinger on tree graph presented in Fig.4 equation (for the case, when the above sum rules are fulfilled) can be shown similarly to that for the star graph. Also, one can show by numerical computations that for the integrable case transmission of nonlocal PT-symmetric solitons are reflectionless. We note that the above treatment can be directly extended to other graph topologies, provided a graph consist of even number number of bonds symmetrically positioned with respect to the origin of coordinates, i.e., one has equal number of bonds on each side of the origin of coordinates. In addition, at least four bonds of the graph should be semi-infinite. Unlike the solution of usual nonlinear Schrodinger equation on graphs, solution of PT-symmetric nonlocal nonlinear NLSE on graphs is much complicated that makes the dynamics of nonlocal solitons more rich than that for usual soliton. The latter implies also existence of more tools for tuning the soliton dynamics.

\section{Conclusions}
In this paper we studied dynamics of solitons described by PT-symmetric nonlocal nonlinear Schrodinger equation on networks by modeling these latter in terms of metric graphs. Integrability of the problem in case of fulfilling certain constraints given in terms of nonlinearity coefficients is shown. Exact soliton solutions which are valid for this case are obtained. For the case, when the constraints are broken, the problem is solved numerically. The analysis of soliton dynamics shows absence of the back scattering in the transmission of soliton through the graph node, is sum rule in Eq. \re{constrain1} is fulfilled. When this sum rule is broken, the transmission is accompanied by scattering of solitons at the node. The treatment is extended for tree graph and possibility for extension for other complicated graphs is discussed. The above model can be applied for describing the soliton dynamics in optical fiber network, where each branch has self-induced gain-loss and other branched waveguides generating self-induced PT-symmetric potential.

\end{document}